\documentclass[twocolumn,prl,aps,superscriptaddress]{revtex4-1}
\usepackage[latin9]{inputenc}
\usepackage{mathtools}
\usepackage{amsbsy}
\usepackage{amstext}
\usepackage{appendix}
\usepackage{braket}
\usepackage{multirow}
\usepackage{array}
\usepackage[unicode=true,pdfusetitle,
 bookmarks=false,
 breaklinks=false,pdfborder={0 0 1},backref=false,colorlinks=false]
 {hyperref}
\usepackage[table]{xcolor}
\usepackage{changepage}
\usepackage{dsfont}
\usepackage{colortbl}
\usepackage{tikz}

\usepackage{multirow}

\definecolor{lightgray}{gray}{0.9}

\usepackage[english]{babel}

\hypersetup{
 bookmarksnumbered=false,bookmarksopen=false}
\makeatletter

\usepackage{amsmath}
\usepackage{graphicx}
\usepackage{amssymb}
\usepackage{txfonts,color}
\@ifundefined{textcolor}{}

{%
 \definecolor{BLACK}{gray}{0}
 \definecolor{WHITE}{gray}{1}
 \definecolor{RED}{rgb}{1,0,0}
 \definecolor{GREEN}{rgb}{0,1,0}
 \definecolor{BLUE}{rgb}{0,0,1}
 \definecolor{CYAN}{cmyk}{1,0,0,0}
 \definecolor{MAGENTA}{cmyk}{0,1,0,0}
 \definecolor{YELLOW}{cmyk}{0,0,1,0}
 \definecolor{ORANGE}{rgb}{1,0.5,0}
}

\newcommand*\circled[1]{\tikz[baseline=(char.base)]{
            \node[shape=circle,draw,inner sep=0.5pt] (char) {#1};}}

\makeatother

\begin{document}
\author{P. Alsina-Bol\'{i}var}
\affiliation{Department of Physical Chemistry, University of the Basque Country UPV/EHU, Apartado 644, 48080 Bilbao, Spain}
\affiliation{EHU Quantum Center, University of the Basque Country UPV/EHU, Leioa, Spain}
\author{J. Casanova}
\affiliation{Department of Physical Chemistry, University of the Basque Country UPV/EHU, Apartado 644, 48080 Bilbao, Spain}
\affiliation{EHU Quantum Center, University of the Basque Country UPV/EHU, Leioa, Spain}

\title{Enhanced microscale NMR spectroscopy of low-gyromagnetic ratio nuclei via hydrogen transfer}

\begin{abstract}
Chemical shifts and J-couplings are fundamental parameters in NMR spectroscopy as they provide structural information about molecules. Extracting these quantities from isotopes such as carbon or nitrogen results in reduced sensitivity due to their low gyromagnetic ratios. In this work, we present a method for detecting chemical shifts and J-couplings at the microscale in low-gyromagnetic-ratio nuclei using NV centers. By leveraging hydrogen nuclei, we achieve strong coupling with NVs and fast signal emission that aligns with NV coherence times. In addition, the technique is well-suited for implementations under high magnetic fields. We demonstrate that our protocol achieves a sensitivity enhancement of more than one order of magnitude for scenarios involving $^{13}$C, with even greater improvements for nuclei with lower gyromagnetic ratios.

\end{abstract}
\maketitle

\section{Introduction}

Nuclear Magnetic Resonance (NMR) is an indispensable tool for elucidating molecular identity, structure, and conformation in a wide range of compounds that comprise organic, inorganic, organometallic, and biological~\cite{cremer2007calculation,roberts2013encyclopedia,pregosin2013nmr,levitt2013spin} molecules. Usually, the nuclei used to perform NMR experiments are hydrogens, both because of their high gyromagnetic ratio and  their large natural abundancy in most molecules (specially in the organic and biological cases).

In the liquid state case, NMR spectroscopy targets, basically, two main intramolecular interactions. These are: (i) Chemical shifts, that reveal the chemical environment around certain nucleus in a molecule, acting as fingerprints to identify specific chemical groups in the compound. (ii) J-couplings, which carry information about molecular structure and conformation through  bond connectivity, as well as of the angle and distance between nuclei~\cite{contreras2000angular,smith1991analysis,schmidt2010correlation,schmidt2011one}. Measuring these target quantities for a specific isotope requires using that nucleus as the emitter. For example, estimating the J-coupling between hydrogen and a $^{31}$P nucleus in a molecule involves tracking oscillations in the NMR signal emitted by hydrogen \cite{glenn2018high}. In cases where the interaction of interest does not involve nuclei with a large gyromagnetic ratio (such as hydrogen or fluorine), the experiment's sensitivity decreases. In addition, sensitivity is further compromised by the low natural abundance of the target nuclei. For example, $^{13}$C is significantly less common than $^1$H, leading to lower amplitudes in $^{13}$C NMR compared to $^1$H NMR \cite{balci2005basic}. n conventional NMR, techniques such as Insensitive Nucleus Enhancement by Polarization Transfer (INEPT) \cite{morris1979enhancement} are commonly used to transfer the higher thermal polarization of hydrogen nuclei to less sensitive nuclei like $^{13}$C, thereby enhancing their polarization and improving detection sensitivity. These methods are also applicable in microscale NMR using NV centers, as they depend only on applying control pulses to the nuclear sample.
Another approach is to increase the static magnetic field and operate in a high-field regime.
Under such conditions, the thermal polarization of nuclei increases, enhancing the amplitude of the emitted NMR signal while  molecular spectra are simplified~\cite{munuera2023high,alsina2024jcoupling}. 

In microscale NMR, nitrogen vacancy (NV) center ensembles have gained significant attention for their ability to probe low-volume samples, offering potential applications in fields such as high-throughput chemistry \cite{mennen2019the}. In this scenario, various protocols including quantum heterodyne measurements with large frequency resolution (up to $\sim 1$ Hz) have been reported~\cite{glenn2018high,bucher2020hyperpolarization,arunkumar2021micron,bruckmaier2023imaging}. In addition, recently introduced protocols have garnered attention for their ability to operate at high magnetic fields on the microscale via NV ensembles~\cite{munuera2023high,daly2024nutation,alsina2024jcoupling,munuera2024high}. This is achieved through an external RF drive applied to the target sample, which induces rotations over nuclei and causes the emission of a slowly oscillating NMR signal which can be tracked by the NV ensemble.
The frequency such NMR signal is constrained by the available RF power as well as by the nature of the targeted nuclei. Consequently, studying nuclei other than hydrogen or fluorine (which have lower gyromagnetic ratios, resulting into slower Rabi frequencies of the applied RF) requires longer detection times. This leads to a significant reduction in sensitivity as a result of sensor decoherence.

In this work, we present a pulse-sequence for performing NMR spectroscopy with NV ensembles over low-gyromagnetic-ratio nuclei by using hydrogens as enhancers of the NMR signal. This approach exploits their high abundance in most molecules and large gyromagnetic ratio, which leads to higher initial polarization and higher coupling with the NV -resulting in fast emission, at a rate matching NV coherence times. Thus, our method not only leverages the high initial polarization of hydrogen nuclei, as in INEPT-like techniques in standard NMR, but also exploits their stronger coupling with the NV ensemble compared to other nuclei. This results in a faster and more intense emitted NMR signal, reducing the detrimental effects of NV decoherence during the measurement period. Additionally, the typically higher concentration of hydrogen relative to other species further enhances the amplitude of the emitted NMR signal. Moreover, the potential for hydrogen hyperpolarization further motivates this approach~\cite{bucher2020hyperpolarization,arunkumar2021micron}. The method is suited for the context of microscale samples analyzed with NV ensembles at high magnetic fields, finally resulting in increased sensitivities.  The protocol can be employed to detect J-couplings, with a resolution limited by the sample's $T_2$, or to detect  J-couplings and chemical shifts, where the resolution is instead limited by $T_2^*$. Through detailed numerical simulations, we demonstrate that our approach improves sensitivity by more than an order of magnitude compared to the only existing standard methods in the scenario of high magnetic fields \cite{munuera2023high,alsina2024jcoupling}.

\section{System}
The Hamiltonian of a given molecule within a liquid reads
\begin{align}
H/\hbar &=\omega^H\sum_i^{N_H}\:S_i^z+\sum_i^{N_H}\delta_i^H\:S_i^z+\sum_j^{N}\omega_j\:I_j^z+\sum_j^{N}\delta_j\:I_j^z\nonumber\\
&+\: \sum_{i<j}^{N_H}{\rm J}_{i,j}^{\rm H}\:\vec{S}_i\cdot\vec{ S}_j +\sum_{i}^{\rm{N_H}}\sum_{j}^{\rm{N}}{\rm{J}}_{i,j}^{\rm het}\:\vec{S}_i\cdot\vec{I}_j+ \sum_{i<j}^{N}{\rm J}_{i,j}\:\vec{I}_i \cdot\vec{I}_j\nonumber\\
&+H_c.
\label{eq:Hamiltonian1}
\end{align}
Note we consider that each molecule contains $N_H$ hydrogens and $N$ non-hydrogen nuclei. 
The first line in Eq.~(\ref{eq:Hamiltonian1}) includes Larmor terms with frequencies $\omega^H$ for hydrogens (with  $S_i=\sigma_i/2$) and their chemical shifts $\delta_i^H$, as well as the corresponding terms for  non-hydrogen nuclei (with $I_j=\sigma_j/2$). The second line encompasses J-couplings among all nuclei. These are: Homonuclear J-couplings among hydrogens (${ J}_{i,j}^{ H}$),  heteronuclear J-couplings (${{J}}_{i,j}^{\rm het}$) that include hydrogen, and J-couplings (both homonuclear and heteronuclear) among non-hydrogen nuclei (${ J}_{i,j}$). 
The last line is the control Hamiltonian, $$H_c=\sum\limits_{i}^{N_H}2\Omega_i\:S_i\cos{\left(\omega_i\:t\right)}+\sum\limits_{i}^{N}2\Omega_i\:I_i\cos{\left(\omega_i\:t\right)}$$ describing the effect of RF fields over the sample. Note that Eq.~\eqref{eq:Hamiltonian1} does not include any term corresponding to the NV sensor, as it only plays a role during the detection of the sample's evolution (see the next section). Additionally, dipole-dipole and other orientation-dependent interactions are neglected, as they average out in liquids.

Equation~(\ref{eq:Hamiltonian1}) is significantly simplifyied using the rotating wave approximation (RWA). This  owes to the difference in Larmor frequencies, specially at the high-field scenario, where $|\omega_j - \omega_i| \gg J_{i,j}, \Omega_i$.  More specifically, in an interaction picture with respect to the Larmor terms, one finds

\begin{align}
H_I/\hbar &=\: \sum_{i<j}^{N_H}{ J}_{i,j}^{ H}\:\vec{S}_i \cdot\vec{S}_j +\sum\limits_{\substack{i<j\\{\rm eq}}}{ J}_{i,j}\:\vec{I}_i \cdot\vec{I}_j+\sum_i^{N_H}\delta_i^H\:S_i^z+\sum_j^{N}\delta_j\:I_j^z\nonumber\\[8pt]
&+\sum_{i}^{{N_H}}\sum_{j}^{{N}}{{J}}_{i,j}^{\rm het}\:S_i^z I_j^z+ \sum\limits_{\substack{i<j\\{\rm neq}}}{ J}_{i,j}\:I_i^z I_j^z\nonumber\\
&+\sum_i^{ N_H}\Omega_i\:S_i^x+\sum_i^N\Omega_i\:I_i^x.
\label{eq:IntHamiltonian1}
\end{align}
Where the labels ``eq" and ``neq" stand for nuclei with equal and not equal Larmor frequencies. Equation~(\ref{eq:IntHamiltonian1}) serves as the starting point for the numerical simulations in Sec.~\ref{sec:numerical}.

\begin{figure*}[]
\includegraphics[width=1 \linewidth]{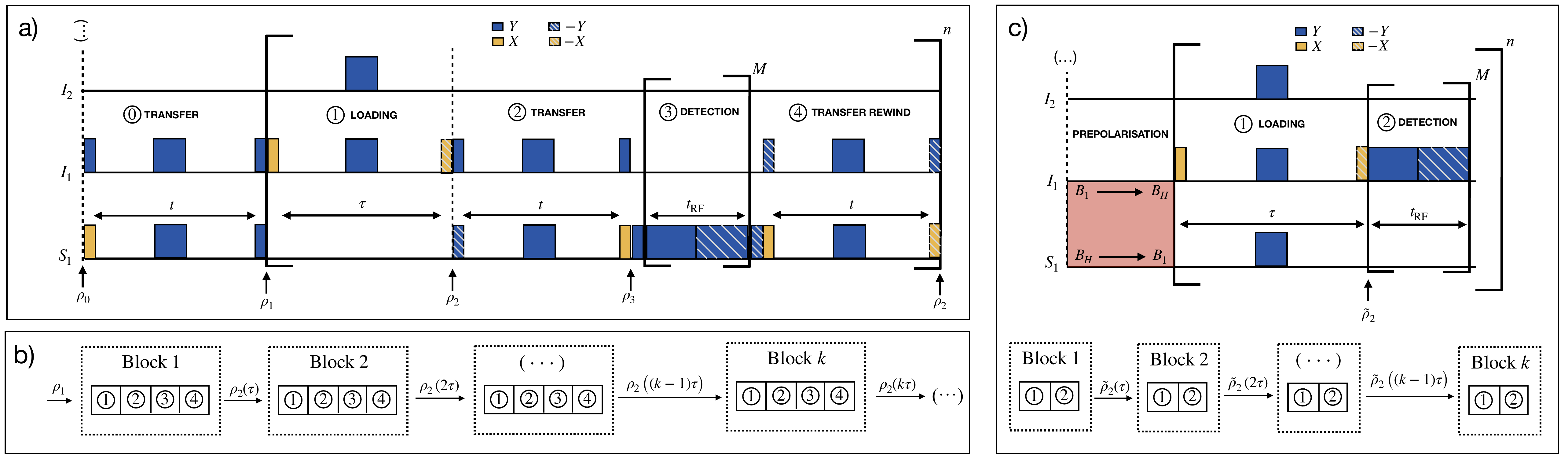}
\caption{\textbf{a)} Schematic representation of our pulse-sequence. The pulse width corresponds to the applied rotation angle (i.e. $\pi/2$ and $\pi$ pulses, as well as the $2\pi$ rotations applied during the detection stages). The $2\pi$ rotations happen only during the detection stages ($\Omega_H\cdot t_{\rm RF}=2\cdot 2\pi$) and are not shown to scale relative to the $\pi/2$ and $\pi$ pulses to avoid excessive spacing in the scheme. Pulse axes follow the color code in the top legend. Horizontal lines correspond to the radiation scheme over  nucleui in the target molecule, namely $S_1$, $I_1$, $I_2$, etc. The main block, repeated $n$ times, is enclosed in brackets (see main text). 
\textbf{b)} Diagram of the input and output density matrices for each block repetition. 
\textbf{c)} (TOP) Standard approach with direct interrogation of nucleus $I_1$ and (BOTTOM) corresponding input/output nuclear state for each block.}

\label{fig:sequence}
\end{figure*}

\section{Protocol}
For the sake of clarity in the presentation of the protocol, we first consider a simple scenario involving a molecule with a single target nucleus, ${I_1}$ (e.g. a $^{13}$C) coupled to a hydrogen, $S_1$, through $J_{1,1}^{\rm het}$ and to other nuclear species (${I_2}$, ${I_3}$ ...) through ${ J}_{1,j}$ ($j>1$). The coupling between $S_1$ and the rest of the nuclei is ${ J}_{1,j}^{\rm het}$ ($j>1$). Our objective is to detect ${ J}_{1,j}$ and $\delta_1$ using the hydrogen ($S_1$) as emitter. 

The Hamiltonian that describes the system is 
\begin{align}
H_I/\hbar =&\delta_{1}^H\:S_i^z+\sum_{j=1}^{N}\delta_j\:I_j^z+\sum_{j=1}^{{N}}{{J}}_{1,j}^{\rm het}\:S_1^z I_j^z+ \sum\limits_{\substack{i<j\\{\rm neq}}}{ J}_{i,j}\:I_i^z I_j^z\nonumber\\[8pt]
&+\Omega_H\:S_1^x+\Omega_1\:I_1^x.
\label{eq:IntHamiltonian2}
\end{align}

Our pulse sequence is depicted in Fig.~\ref{fig:sequence} a). Firstly, a \textit{transfer} stage \circled{0} is applied. Then,  a block consisting of four stages: \circled{1} {\it loading}, \circled{2} {\it transfer}, \circled{3} {\it detection} (repeated $M$ times to optimize sensitivity), and \circled{4} {\it transfer rewind}. This block is repeated $n$ consecutive times  (see details later). 

Now, we analyze the evolution of the whole nuclear state at every stage of the protocol and track the journey of  $\delta_1$ and $J_{1,j}$ from the target nucleus $(I_1)$ to the hydrogen (refer to Appendix~\ref{app:calcs} for general molecules). The initial state of the nuclei, see Fig.~\ref{fig:sequence} a), is

\begin{align}\label{rho0}
    \rho_0=\frac{1}{2^{(N-1)}}\left(\mathds{1}+B_H\:S_1^z+B_1\:I_1^z+\sum_{j=2}^NB_j\:I_j^z\right),
\end{align}
where the Boltzmann factors are  $B_H=\frac{\hbar\:\gamma_H\:B_z}{K_B\:T}$ and $B_j=\frac{\hbar\:\gamma_j\:B_z}{K_B\:T}$, $\gamma_H$ and $\gamma_j$ are the gyromagnetic ratios of the hydrogen and nucleus ${I_j}$ respectively. $B_z$ Is the static magnetic field in the $z$-direction, $K_B$ the Boltzmann constant, and $T$ is the temperature. Note that, in Eq.~(\ref{rho0}) we only consider first order terms in the Boltzmann factors since the contribution of higher orders is negligible.

In the first \textit{transfer} stage \circled{0}, see Fig.~\ref{fig:sequence} a), $S_1$ and $I_1$ become correlated through the interaction $J_{1,1}^{\rm het}$. Note that the simultaneous application of $\pi$-pulses refocuses all the terms in Eq.~(\ref{eq:IntHamiltonian2}) related to $S_1$ and $I_1$, except $J_{1,1}^{\rm het}\:S_1^z\:I_1^z$. In addition, the delivered $\pi$-pulses would reduce inhomogeneous broadening. 

For a time $t=\frac{1}{2\:J_{1,1}^{\rm het}}$, $\rho_0$ evolves to 

\begin{align}
    \rho_1=\frac{1}{2^{(N+1)}}\left[\mathds{1}_1-8\:B_H\:S_1^z\:I_1^x+8\:B_c\:S_1^x\:I_1^y+\sum_{j=2}^N\:B_j\:I_j^z\right].  \nonumber
\end{align}

Now, the block depicted in Fig.~\ref{fig:sequence} a) begins. In the \textit{loading} stage \circled{1}, the $\pi$-pulses applied to all nuclei $I_i$ refocus every interaction in Eq.~(\ref{eq:IntHamiltonian2}) except for $ \sum\limits_{\substack{i<j\\{\rm neq}}}{ J}_{i,j}\:I_i^z I_j^z$. Note that if these $\pi$-pulses are not delivered, while we decouple $S_1$ from the rest of spins (via, e.g., a midway $\pi$-pulse in the $S_1$ channel), the non-refocused interactions are $ \sum\limits_{\substack{i<j\\{\rm neq}}}{ J}_{i,j}\:I_i^z I_j^z + \sum\limits_{j=1}^{N}\delta_j\:I_j^z$. In either case, the nuclear state after $\circled{1}$ is

\begin{align}
\rho_2(\tau)=\frac{1}{8}&\left\{\mathds{1}-8\:B_H\:S_1^z\:I_1^x\:C(\tau)+8\:B_c\:S_1^x\:I_1^y+2\sum_{j=2}^N\:B_j\:I_j^z\right\},\label{eq:state_tau}
\end{align} 
where the term involving the target quantities is $C(\tau) = \prod\limits_{j=2}^{N} \left[ \cos{\left( \frac{J_{1,j}}{2} \tau \right)} \right]$ or $C(\tau) = \prod\limits_{j=2}^{N} \left[ \cos{\left( \frac{J_{1,j}}{2} \tau \right)} \right] \cos{\left( \delta_1 \tau \right)}$, depending on whether the $\pi$-pulses are applied, or not. Consequently, the coherence time of the nucleus $I_1$ is $T_2$, or $T_2^*$. 
For simplicity, only the terms from the evolution during \textit{loading} stage  \circled{1} that contribute to the emitted NMR signal are retained in Eq.~(\ref{eq:state_tau}).

The next \textit{transfer} stage \circled{2} encodes $C(\tau)$ in the population of $S_1$ such that

\begin{align}
\rho_3(\tau)=\frac{1}{8}&\left\{\mathds{1}-2\:B_H\:S_1^z\:C(\tau)+2\:B_c\:I_1^z+2\sum_{j=2}^N\:B_j\:I_j^z\right\},
\end{align}
with the expectation value $\langle S_1^z\rangle$ of the hydrogen nucleus containing $C(\tau)$ as

\begin{align}
    \langle S_1^z\rangle(\tau)=-\frac{1}{2}B_H\:C(\tau).
    \label{eq:expect}
\end{align}

At the \textit{detection} stage \circled{3}, a number $M$ of $2\pi$ and $-2\pi$ rotations along the $y$-axis are applied to the hydrogen nucleus, each with a duration $t_{\rm RF}$. 
Due to this rotations and to the scenario we consider -NV ensemble at a sufficient distance from the sample- the interaction between the sample and the NV ensemble (dipole-dipole) can be can be described by a classical NMR signal emitted from the sample  $B_z(t)$~\cite{meriles2010imaging}, with the form

\begin{equation}\label{eq:effectiveB}
B_z(t)\propto \gamma_H\langle S_1^z \rangle(\tau)\sin(\Omega^\text{H} \:t).
\end{equation}
where $\Omega^\text{H}\propto\gamma_H$ is the RF Rabi frequency, which coincides with the speed of the NMR signal. We remark that $B_z(t)$ encodes $C(\tau)$ --thus $\delta_1$ and $J_{1,j}$-- according to Eq.~(\ref{eq:expect}). Refer to Appendix \ref{app:readout} for further details. $B_z(t)$ is subsequently measured by the NV ensemble, allowing for the recovery of $\delta_1$ and $J_{1,j}$ (as discussed below). Note that the NMR signal frequency, $\Omega^\text{H}$, can be tuned, for instance to tens of kHz~\cite{herb2020broadband, yudilevich2023coherent}, such that it is easily tracked by the NV ensemble at high fields. 

Remarkably, using hydrogen as emitters at this stage offers two key advantages: First, the larger thermal polarization of hydrogen nuclei (note this is a treat widely used in standard NMR~\cite{morris1979enhancement}) and, second, the rapid spinning of hydrogen under RF radiation allows their oscillations to be tracked by NVs within their coherence time.  

In this stage, $M$ represents the number of measurements performed by the NV ensemble (we discuss the optimal value for $M$ in Sec.~\ref{sec:sens}). Finally, we note that when completing $M$ rotations, the nuclear state remains at $\rho_3(\tau)$, while the $2\pi$ and $-2\pi$ precessions are introduced for robustness~\cite{munuera2023high, daly2024nutation, alsina2024jcoupling}.  Following the \textit{detection} stage, a \textit{transfer rewind} \circled{4} is implemented, returning the nuclear state to $\rho_2(\tau)$. Now, the entire block (comprising stages \circled{1}, \circled{2}, \circled{3}, and \circled{4}) is applied again. After  the new \textit{loading} stage \circled{1} we get $\rho_2(2\tau)$. 

Generally, at the $k$-th block, the nuclear state entering the {\it transfer} stage \circled{2} is $\rho_2(k\tau)$, see Fig.~\ref{fig:sequence} b), thus the NMR signal at the corresponding \textit{detection} stage is 

\begin{equation}\label{eq:effectiveBn}
B_z(t)\propto \gamma_H\langle S_1^z \rangle(k\tau)\sin(\Omega^\text{H} \:t),
\end{equation}
with
\begin{align}
    \langle S_1^z\rangle(k\tau)=-\frac{1}{2}B_H\:C(k\tau),
    \label{eq:expectn}
\end{align}
and  $C(k\tau)$~is~$\prod\limits_{j=2}^N\left[\cos{\left(\frac{J_{1,j}}{2}k \tau\right)}\right]$ or $\prod\limits_{j=2}^N\left[\cos{\left(\frac{J_{1,j}}{2}k\tau\right)}\right]\:\cos{\left(\delta_1k\tau\right)}$ depending if $\pi$-pulses are applied or not.

One can demonstrate that, at the $k$-th block (see Appendix~\ref{app:readout}) the NV expected value reads
\begin{equation}\label{eq:exp_value_nv}
\langle\sigma_y\rangle^{\rm NV}\propto \langle S_1^z \rangle(k\tau).
\end{equation}

\section{Sensitivity}\label{sec:sens}
We compare the performance of our protocol with respect to a standard approaches \cite{munuera2023high,alsina2024jcoupling} where the $I_1$ nucleus is directly interrogated, see Fig.~\ref{fig:sequence} c).  Note that the experimental setup is the same in all cases: a microscale sample positioned above an NV ensemble that detects its emitted magnetic field. The only thing that changes is the control RF applied to the sample.

One can check that, at the $k$-th detection stage of the standard approach, we have  (see Appendix~\ref{app:simpler})

\begin{equation}\label{eq:effectiveB1}
B_z(t)\propto \gamma_1\langle I_1^z \rangle(k\tau)\sin(\Omega^\text{1} \:t),
\end{equation}
where, importantly, the amplitude  is $\propto \gamma_1\langle I_1^z \rangle(k\tau)$, the frequency $\Omega^1\propto\gamma_1$, and

\begin{align}
    \langle I_1^z\rangle(k\tau)=-\frac{1}{2}B_H\:C(k\tau).
    \label{eq:expectn_i}
\end{align}

To ensure a fair comparison with our protocol, we assume in Eq.~(\ref{eq:expectn_i}) that nucleus $I_1$ has the same expectation value as hydrogen, as indicated by the presence of $B_H$ in Eq.~(\ref{eq:expectn_i}). This can be achieved through a prepolarization stage, see Fig.~\ref{fig:sequence}~c). Otherwise, without this assumption, the results would be even more favorable for our protocol.

We have developed a sensitivity ratio  (see Appendix~\ref{app:sensitivity})  between our proposal, and the standard approach. This is 
\begin{align}
    \frac{\eta_{\rm NV}^H}{\eta_{\rm NV}^{1}}=\left[\frac{1-e^{-\frac{\tau}{T_{2, \rm eff}^1}n}}{1-e^{-\frac{\tau}{T_{2, \rm eff}^H}n}} \right]\frac{T_{2,\rm eff}^1}{T_{2,\rm eff}^H}\: \frac{A_1}{A_H}\sqrt{\frac{2\:t+\tau+M\cdot t_H}{\tau+M_1\cdot t_1}}\sqrt{\frac{M_1}{M}},
    \label{eq:sens_general}
\end{align}
where $\eta_{\rm NV}^H$ ($\eta_{\rm NV}^1$) is the sensitivity of our method (the standard one).
Equation~(\ref{eq:sens_general}) serves as a valuable metric for assessing whether our approach outperforms the standard case. Specifically, when  $ \frac{\eta_{\rm NV}^H}{\eta_{\rm NV}^{1}} < 1$ it indicates that our proposal offers improved sensitivity.

Further parameters in Eq.~(\ref{eq:sens_general}) are: The effective decoherence times of the hydrogen and the $I_1$ nucleus, $T_{\rm 2, eff}^H$ and $T_{\rm 2, eff}^1$, (see Appendix~\ref{app:sensitivity} for explicit expressions). The parameters $n$, $\tau$, and $t$ are time intervals defined in Fig.~\ref{fig:sequence} a), while $t_H$ and $t_1$ represent the total time that hydrogen and $I_1$ nuclei are exposed to radiation. $M$ and $M_1$ represent the number of \textit{detection} stages per block in our sequence and the standard approach. Finally, the functions $A_1$ and $A_H$ in Eq.~(\ref{eq:sens_general}) describe the accumulated phase by the NV ensemble in each method during each detection stage (see Appendix~\ref{app:readout}). These are

\begin{align}
    A_H \propto \frac{2\gamma_e\:\gamma_H}{\Omega^H} \langle S_1^z \rangle(k\tau) \left[1 - \cos{\left(\pi\:T_2^{\rm NV}\Omega^H\right)}\right],
    \label{eq:NVphaseH}
\end{align}
for hydrogen, where $\gamma_e$ and $\gamma_H$ are the gyromagnetic ratios of the electron and hydrogen, respectively, $\Omega^H$ is the Rabi frequency of the applied RF. Similarly, interrogating nucleus $I_1$ leads to

\begin{align}
    A_1 \propto \frac{2\gamma_e\:\gamma_1}{\Omega^1} \langle I_1^z \rangle(k\tau) \left[1 - \cos{\left(\pi\:T_2^{\rm NV}\Omega^1\right)}\right],
    \label{eq:NVphaseI1}
\end{align}
where $\gamma_1$ is the gyromagnetic ratio of $I_1$, $\Omega^1$ is the Rabi frequency of the applied RF.

The term $\frac{A_1}{A_H}$ in Eq.~(\ref{eq:sens_general}) is key to our approach, as nuclei with lower gyromagnetic ratios than hydrogen result in less phase accumulation by the NV ensemble contributing to achieve  $ \frac{\eta_{\rm NV}^H}{\eta_{\rm NV}^{1}} < 1$. In addition, the factor $\sqrt{\frac{M_1}{M_H}}$ further benefits our method, since $\gamma_H > \gamma_1$ implies $M_H > M_1$.

To finish, note that the sensitivity improvement comes at the cost of reduced resolution, as our method shortens the effective decoherence times of the nuclei in the sample (see Appendix \ref{app:sensitivity} for explicit expressions).

\section{Numerical Results}\label{sec:numerical}

We begin by demonstrating the validity of the protocol with a simple case: a hydrogen cyanide molecule --${\rm ^1H ^{13}C ^{15}N}$-- containing a single hydrogen atom. Now we list the relevant parameter used across this section:
The couplings are \cite{he2012structural} ${ J_{H-C}}=\left(2\pi\right)\times\left(267\right){\rm Hz}$, ${ J_{H-N}}=\left(2\pi\right)\times\left(-11\right){\rm Hz}$  and ${ J_{C-N}}=\left(2\pi\right)\times\left(-25\right){\rm Hz}$, which leads to $t=\frac{1}{2\:J_{H-C}}=1.9$ ms. The external magnetic field is  set to 2 T and room temperature is assumed ($\sim 300$ K). The corresponding gyromagnetic ratios are $\gamma_{^1 H}=(2\pi)\times42.6\: {\rm MHz\cdot T^{-1}}$, $\gamma_{^{13}C}=(2\pi)\times10.7\: {\rm MHz\cdot T^{-1}}$ and $\gamma_{^{15}N}=(2\pi)\times-4.3\: {\rm MHz\cdot T^{-1}}$, which result in the Boltzmann factors $B_{\rm H}\sim1.4\cdot 10^{-5}$, $B_{\rm C}\sim 3.5\cdot 10^{-6}$ and $B_{\rm N}\sim 1.4\cdot 10^{-6}$. Dephasing noise of every specie was implemented via an exponential decay where: $T_2^{H}=1$ s, $T_2^{C}=4$ s, $T_2^{C*}=0.4$ s  $T_2^{N}=6$ s and $T_2^{\rm NV}=10\:\mu$s. 
In both approaches in Figs.~\ref{fig:sequence} a) and c), we set $\tau = 1$ ms, and $n = 240$ resulting in a total time of $\tau^{\rm tot} = 0.24$~s for both. 

Regarding the sensor, we assume an NV ensemble at $1\mu$m depth, leading to a detectable sample in the order of pL \cite{glenn2018high,meriles2010imaging}. The NV concentration is approximately $2\cdot10^{17} {\rm cm}^{-3}$, with a total active NV volume of $V=\pi\times\left(5\cdot10^{-5} \right)^2\times 13\cdot 10^{-6} =10^{-15}{\rm m}^{-3}$, all extracted from \cite{glenn2018high}. This results in an estimated total of $\sim10^8$ NV centers.

\begin{figure}[t!]
\includegraphics[width=1 \linewidth]{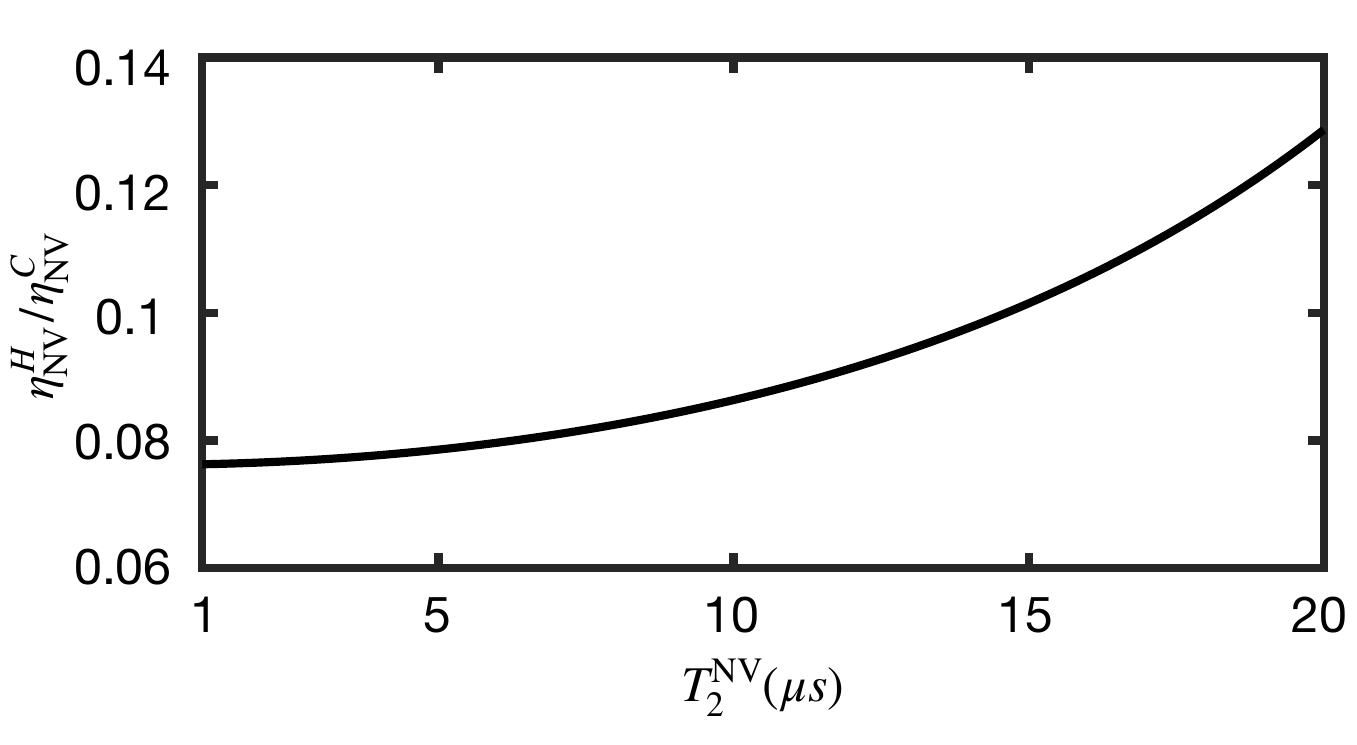}
\caption{Sensitivity ratio between our approach and the standard one for the ${\rm ^1H ^{13}C ^{15}N}$ molecule as a function of $T_2^{\rm NV}$.}
\label{fig:comp}
\end{figure}

\subsection{J-coupling detection}
Here we utilize our method in Fig.~\ref{fig:sequence}~a) including midway $\pi$-pulses in the \textit{loading} stage \circled{1} for J-coupling detection. Under these conditions Eq.(\ref{eq:expectn}) simplifies to

\begin{figure*}[]
\includegraphics[width=0.8 \linewidth]{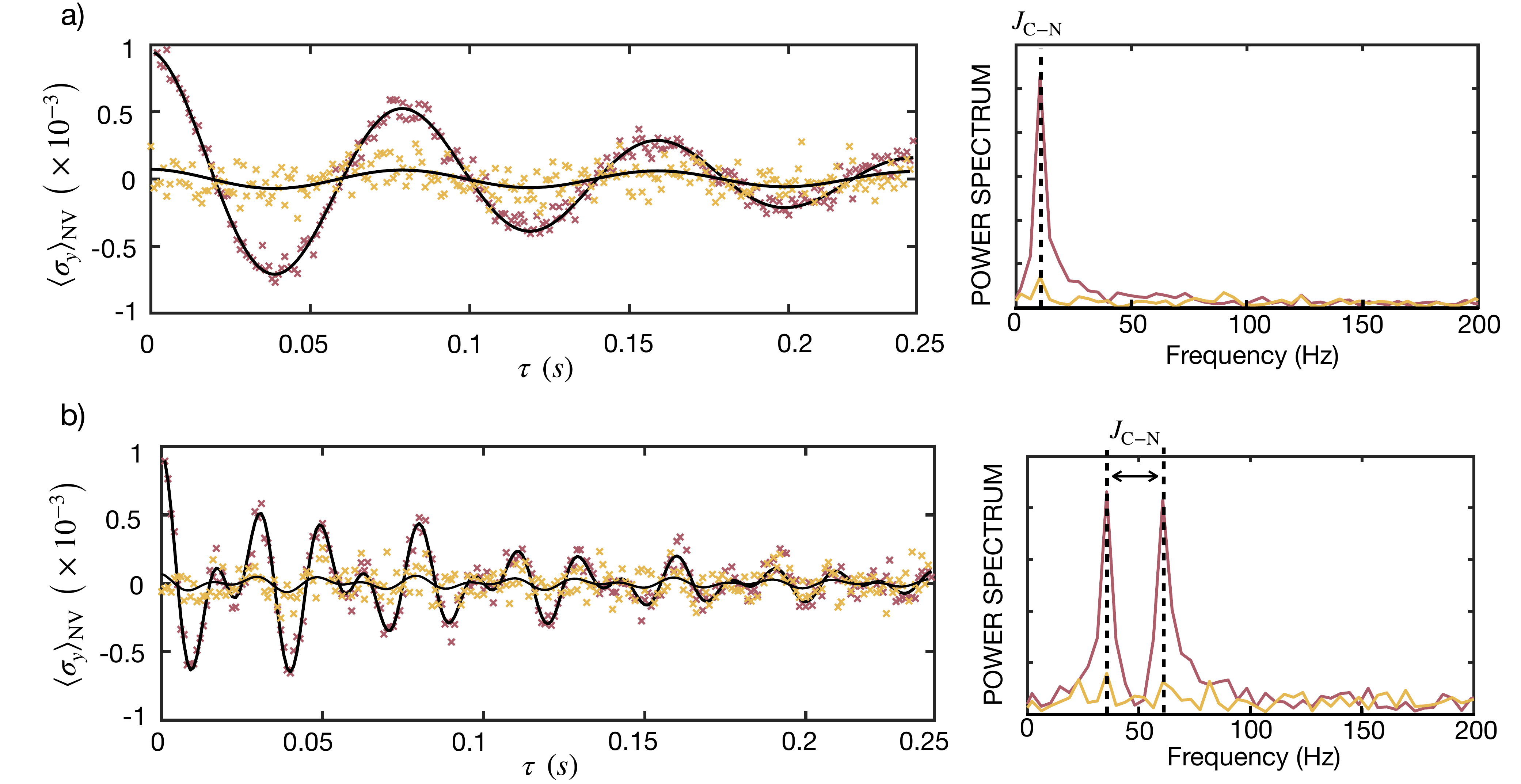}
\caption{Expected value of the NV ensemble signal (left) and its Fourier transform spectra (right) are shown for the ${\rm ^1H ^{13}C ^{15}N}$ molecule. Red data correspond to our method, while yellow to the standard approach with a direct interrogation of $^{13}$C (solid black lines fits the obtained data). The total experimental time is \( t_{\rm tot} = 10^3 \) s for both cases, while simulations account for NV fluorescence with a 7\% contrast (as reported in~\cite{glenn2018high}). 
\textbf{(a)} Both sequences include $\pi$-pulses during the \textit{loading} stages. The red and yellow peaks in the spectrum are centered at the same frequency, while  our method results in a higher SNR.
\textbf{(b)} Midway $\pi$-pulses are removed during the \textit{loading} stages, allowing chemical shift detection. The peaks are now centered at the chemical shift value, with a splitting due to the J-coupling between carbon and nitrogen.
}
\label{fig:results_HCN}
\end{figure*}
\begin{align}
        \langle S_1^z\rangle(k\tau)=-\frac{1}{2}B_H\cos\left(\frac{J_{\rm C-N}}{2}\:k\tau\right),
        \label{eq:exp_hcn}
\end{align}
Where we relabelled $J_{H,1}^{\rm het}\rightarrow J_{\rm H-C}$ and $J_{1,2}\rightarrow J_{\rm C-N}$ for simplicity.
The objective is to read $J_{\rm C-N}$ via hydrogen and compare its performance with the standard approach in Fig.~\ref{fig:sequence}~ c). 

Before presenting the numerical results we make some estimations using Eq.~(\ref{eq:sens_general}). In particular,  with  the optimal number of measurements  $M=70$ and $M_1=19$ we get 
\begin{align}
    \frac{\eta_{\rm NV}^H}{\eta_{\rm NV}^{C}}\approx 0.09,\nonumber
\end{align}
which indicates the  $ \frac{1}{0.09}\approx 11.14$ times better sensitivity exhibited by of our method. 
This performance is strongly dependent on the decoherence time of the NV ensemble ($T_2^{\rm NV}$) as it affects  $A_H$ and $A_1$, see Eqs.~(\ref{eq:NVphaseH}) and (\ref{eq:NVphaseI1}). In Fig.~\ref{fig:comp}, we investigate the ratio $  \frac{\eta_{\rm NV}^H}{\eta_{\rm NV}^{C}}$ for distinct values of $T_2^{\rm NV}$. Notably, for $T_2^{\rm NV}$ ranging from one microsecond to tens of microseconds, our protocol demonstrates improved sensitivity.

Regarding numerical results, Fig.~\ref{fig:results_HCN}~a) depicts the expected value of the NV ensemble $(\langle\sigma_{y}\rangle_{\rm NV})$ reconstructed from their photoluminescence (left) and the corresponding spectrum (right). Red data correspond to our proposal in Fig.~\ref{fig:sequence}~a) and yellow to the standard approach in Fig.~\ref{fig:sequence} c). 
The signal-to-noise ratio (SNR) in our case is significantly enhanced due to the higher amplitude of the NMR signal achieved via hydrogen transfer.
More specifically, we get $\frac{\rm SNR_H}{\rm SNR_C} = \left(\frac{\eta_{\rm NV}^H}{\eta_{\rm NV}^{C}}\right)^{-1}= 11.1$ (where $\rm SNR_H$ and $\rm SNR_C$ are the signal to noise ratios of our method and the standard approach) which demonstrates the superior performance of our protocol.

\subsection{Chemical shifts and J-couplings detection}\label{sec:cs}

To detect chemical shifts, $\pi$-pulses during the \textit{loading} stages \circled{1} are removed. In this scenario,  inhomogeneous broadening over hydrogen is not cancelled leading to reduced spectral resolution  (note, in this case the effective decoherence time is  $T_2^{C*}$ instead of $T_2^{C}$). 

Following Eq.~(\ref{eq:expectn}) and (\ref{eq:expectn_i}), we find
\begin{align}
 \langle S_1^z\rangle= \langle I_1^z\rangle=-\frac{1}{2}B_H\cos\left(\delta_{C}\:k\tau\right)\cos\left(\frac{J_{C-N}}{2}k\tau\right).\nonumber
\end{align}
Figure~\ref{fig:results_HCN}~b) shows the obtained NV expectation value (left) and spectra (right) for this scenario. The spectrum is now centered at the chemical shift value, taken as $\delta_C =(2\pi)\times 50$ Hz, while the J-coupling causes the splitting into two peaks. 

Table~\ref{table:values} shows values for the ratio $\frac{\rm SNR_H}{\rm SNR_C}$ obtained via 
Eq.~(\ref{eq:sens_general}), $\left(  \frac{\rm SNR_H}{\rm SNR_C}\right)_{\rm eq} = 11.14$, and from numerical simulations, $\left(\frac{\rm SNR_H}{\rm SNR_C}\right)_{\rm num}=11.10$. Note the agreement between both values, and the superior performance of our method as the obtained ratio is larger than one.

\subsection{Complex molecules}\label{sec:complex}

In the general case encompassing with $N_H$ hydrogens, $N_t$ target nuclei, and $N$ other nuclei, one can demonstrate (see Appendix \ref{app:calcs}) 

\begin{align}
    \langle S_i^z\rangle=&B_H\sum_j^{N_t}\left\{\sin^2{\left(\frac{J_{i,j}^{\rm het}}{2}t\right)}\prod_{k\neq j}^{N_t}\left[\cos^2{\left(\frac{J_{i,k}^{\rm het}}{2}t\right)}\right]\prod_{k\neq j}^{N}\left[\cos{\left(\frac{J_{j,k}}{2}\tau\right)}\right]\right\},
        \label{eq:exp_z_complicated}
\end{align}
 with $\langle S^z\rangle=\sum\limits_{i}^{N_H}\langle S_i^z\rangle$.

We present now the results obtained applying our method and the standard approach to a trimethylphosphine molecule, $\rm{P\left(CH_3\right)_3}$. Owing to computational restrictions,  we consider a scenario where only one of the three carbons is a $^{13}\text{C}$ (i.e., a spin 1/2 particle). Heteronuclear J-couplings are $J_{H_1-C}=(2\pi)\times103\:\rm{Hz}$, $J_{H_2-C}=\left(2\pi\right)\times\:80$ Hz and $J_{C-P}=\left( 2\pi\right)\times\:47$ Hz, while the homonuclear ones  are $J_{i,j}^H = (2\pi)\times20$ Hz for hydrogens within the same methyl group, and $J_{i,j}^H = (2\pi)\times8$ Hz for hydrogens in different ones.
Finally, the \textit{transfer} stage duration is set to $t=5.7$ ms.

\setlength{\tabcolsep}{12pt} 
\renewcommand{\arraystretch}{1.5} 
\begin{table*}[]
\begin{tabular}{ | c | c | c | c |c | } 
  \hline
   & \multicolumn{2}{c|}{\bf{J-Couplings}} & \multicolumn{2}{c|}{\bf{J-Couplings and Chemical Shifts}} \\ 
  \hline
   \bf{Molecule} & $\left(  \frac{\rm SNR_H}{\rm SNR_C}\right)_{\rm eq}$ & $\left(  \frac{\rm SNR_H}{\rm SNR_C}\right)_{\rm num}$ & $\left(  \frac{\rm SNR_H}{\rm SNR_C}\right)_{\rm eq}$ & $\left(  \frac{\rm SNR_H}{\rm SNR_C}\right)_{\rm num}$ \\ 
  \hline
   ${\rm ^1H ^{13}C ^{15}N}$ & $11.14$ & $11.10$ & $11.8$ & $11.15$ \\ 
  \hline
  \bfseries $\rm P(CH_3)_3$ & 45.5 & 33.5 & 71.0 & 53.1 \\ 
  \hline
\end{tabular}
\caption{Values for $\frac{\rm SNR_H}{\rm SNR_C}$ calculated from the theoretical expression, $\left(  \frac{\rm SNR_H}{\rm SNR_C}\right)_{\rm eq}$, and via numerical simulations, $\left(  \frac{\rm SNR_H}{\rm SNR_C}\right)_{\rm num}$, for the ${\rm ^1H ^{13}C ^{15}N}$ and $\rm P(CH_3)_3$ molecules. }
\label{table:values}
\end{table*}

In terms of sensitivity, we modify Eq.~(\ref{eq:sens_general}) adding the amplitude term coming from Eq.~(\ref{eq:exp_z_complicated}). This results in the following expression

\begin{align}
    \frac{\eta_{\rm NV}^H}{\eta_{\rm NV}^{1}}=&\left[\frac{1-e^{-\frac{\tau}{T_{2, \rm eff}^1}n}}{1-e^{-\frac{\tau}{T_{2, \rm eff}^H}n}} \right]\frac{T_{2,\rm eff}^1}{T_{2,\rm eff}^H}\: \frac{A_1}{A_H}\sqrt{\frac{2\:t+\tau+M\cdot t_H}{\tau+M_1\cdot t_1}}\sqrt{\frac{M_1}{M}}\nonumber\\[8pt]
   & \times\frac{1}{3\sin^2\left(\frac{J_{H_1-C}^{\rm het}}{2}t\right)+6\sin^2\left(\frac{J_{H_2-C}^{\rm het}}{2}t\right)}.
    \label{eq:sens_general_trimethyl}
\end{align}

In Appendix~\ref{app:calcs}, Fig.~\ref{fig:PCH33_results}~a) illustrates the NV expectation value and the corresponding spectra for our method as well as for the standard approach. As in the previous example, one can choose to either retain or omit the $\pi$-pulses during the \textit{loading} stage \circled{1} in Fig.~\ref{fig:sequence}~a) and Fig.~\ref{fig:sequence}~c) to either suppress or detect chemical shifts. In Table~\ref{table:values}, we present the expected SNR ratio, computed using Eq.~\eqref{eq:sens_general_trimethyl}, i.e. $\left(  \frac{\rm SNR_H}{\rm SNR_C}\right)_{\rm eq}$ alongside the corresponding ratio obtained from numerical simulations $\left(  \frac{\rm SNR_H}{\rm SNR_C}\right)_{\rm num}$. The observed discrepancy between the calculated and simulated SNRs is attributed to the J-couplings among hydrogens (see Eq.~(\ref{eq:IntHamiltonian1}), first line), which are not accounted in the analytical expression. Nonetheless, the results demonstrate a significant improvement in the SNR of our method compared to the standard approach, even for complex molecules.

\section{Conclusions}
We present a protocol that enables the detection of NMR signals encoding parameters of low-gyromagnetic-ratio nuclei (such as $^{13}$C) through hydrogen nuclei.  By leveraging the high gyromagnetic ratio and the common abundance of hydrogen atoms in target molecules, our protocol significantly enhances the signal-to-noise ratio. It achieves more than one order of magnitude improvement for $^{13}$C, while even greater gains are expected for nuclei with lower gyromagnetic ratios.

\section{Acknowledgements } We thank Professor Dominik B. Bucher for his comments on the manuscript. P. A. B. acknowledges support from UPV/EHU Ph.D. Grant No. PIF 23/275. J. C. acknowledges the Ram\'{o}n y Cajal (RYC2018-025197-I) research fellowship. Authors acknowledge the Quench project that has received funding from the European Union's Horizon Europe -- The EU Research and Innovation Programme under grant agreement No 101135742, the Spanish Government via the Nanoscale NMR and complex systems project PID2021-126694NB-C21, and the Basque Government grant IT1470-22.

\onecolumngrid
\appendix
\counterwithin{figure}{section}

\newpage

\section{Analytical calculation of the dynamics}\label{app:calcs}
We provide a detailed calculation of the dynamics experienced by the nuclei at each stage of our sequence in Fig.~\ref{fig:sequence}~a). We label the hydrogen spins as $S_i$, the target nuclei (e.g. $^{13}\text{C}$ nuclei) as $I_i$, and the rest of the nuclei as $\sigma_i$. We consider a molecule with $N_H$ hydrogens, $N_t$ target nuclei, and a number $N$ of other nuclei. 

In the \textit{transfer} stage, the Hamiltonian and the corresponding propagator is

\begin{align}
H_1=\sum_{i=1}^{N_H}\sum_{j=1}^{N_t}J_{i,j}^{\rm het}S_i^z\:I_j^z\rightarrow U_1=e^{-i\:\left(\sum\limits_{i=1}^{N_H}\sum\limits_{j=1}^{N_t}J_{i,j}^{\rm het}S_i^z\:I_j^z\right)\:t}
\label{eq:app:dynt}
\end{align}

In the loading stage we have

\begin{align}
H_2=\sum_{i=1}^{N_t}\sum_{j=1}^{N}J_{i,j}I_i^z\:\sigma_j^z\rightarrow U_2=e^{-i\:\left(\sum\limits_{i=1}^{N_t}\sum\limits_{j=1}^{N}J_{i,j}I_i^z\:\sigma_j^z\right)\:\tau}e^{-i\left(\sum\limits_{i=1}^{N_t}\delta_iI_i^z\right)\:\tau}e^{-i\left(\sum\limits_{i=1}^{N}\delta_iI_i^z\:\right)\tau}.
\label{eq:app:dync}
\end{align}

The initial state is: (note we disregard terms corresponding to  $\sigma_i$ nuclei since they do not contribute to the measured signal)

\begin{align}
\rho_0=B_H\sum_{i=1}^{N_H}S_i^z+B_I\sum_{i=1}^{N_t}I_i^z+\mathds{1}
\label{eq:app:rho_0}
\end{align}

By evolving  $\rho_0$ with the first \textit{transfer} dynamics ($U_1$), we get

\begin{align}
\rho_1=&-B_H\sum_{i=1}^{N_H}S_i^y\prod_{j=1}^{N_t}\left[\cos{\left(\frac{J_{i,j}^{\rm het}}{2}\:t\right)}+i\sin{\left(\frac{J_{i,j}^{\rm het}}{2}\:t\right)}\:S_i^x\:I_j^x\right]\\[8pt]
&-B_I\sum_{i=1}^{N_t}I_i^x\prod_{j=1}^{N_H}\left[\cos{\left(\frac{J_{i,j}^{\rm het}}{2}\:t\right)}+i\sin{\left(\frac{J_{i,j}^{\rm het}}{2}\:t\right)}\:S_i^x\:I_j^x\right]+\mathds{1}
\label{eq:app:rho_1}
\end{align}

Now we only retain terms in the density matrix that contributes to the detectable NMR signal. Specifically, we keep only the first line of the previous equation. Moreover, only the terms with an odd number of sine functions contribute to the signal, as these correspond to the $S_i^z$ coherence of the hydrogen nuclei, which produces the measured signal (see the simple case in the main text). Thus, we can write

\begin{align}
\rho_1\rightarrow&-B_H\sum_{i=1}^{N_H}S_i^z\left[\sum_{j=1}^{N_t}A_j^i\:I_j^x+\sum_{j=1}^{N_t}\sum_{p=1}^{N_t}\sum_{q=1}^{N_t}A_{j,p,q}^i\:I_j^x\:I_p^x\:I_q^x+...\right],
\label{eq:app:rho_1}
\end{align}

with $$A_j^i\equiv \sin{\left(\frac{J_{i,j}^{\rm het}}{2}\:t\right)\prod\limits_{k\neq j}^{N_t}\left[\cos{\left(\frac{J_{i,k}^{\rm het}}{2}\:t\right)}\right]},$$ $$A_{j,p,q}^i\equiv -\sin{\left(\frac{J_{i,j}^{\rm het}}{2}\:t\right)}\sin{\left(\frac{J_{i,p}^{\rm het}}{2}\:t\right)}\sin{\left(\frac{J_{i,q}^{\rm het}}{2}\:t\right)}\prod\limits_{\substack{k\neq j\\k\neq p\\k\neq q}}^{N_t}\left[\cos{\left(\frac{J_{i,k}^{\rm het}}{2}\:t\right)}\right].$$ 

In the loading stage \circled{1} we apply $U_2$. After such evolution, the state reads

\begin{align}
\rho_2\rightarrow&-B_H\sum_{i=1}^{N_H}S_i^z\left[\sum_{j=1}^{N_t}A_j^i\:I_j^xe^{-2i\:\left(\sum\limits_{k=1}^{N}J_{j,k}I_j^y\:\sigma_k^z\right)\:\tau}e^{2i\:\delta_j\:I_j^z\tau} +\sum_{j=1}^{N_t}\sum_{p=1}^{N_t}\sum_{q=1}^{N_t}A_{j,p,q}^i\:I_j^x\:I_p^x\:I_q^xe^{-2i\sum\limits_{k=1}^{N}\left(J_{j,k}I_j^y+J_{p,k}I_p^y+J_{q,k}I_q^y\right)\sigma_k^z\:\tau}e^{2i\left(\:\delta_j\:I_j^z+\delta_p\:I_p^z+\delta_q\:I_q^z\right)\tau}+...\right],
\label{eq:app:rho_2}
\end{align}

\begin{figure*}[t]
\includegraphics[width=0.8 \linewidth]{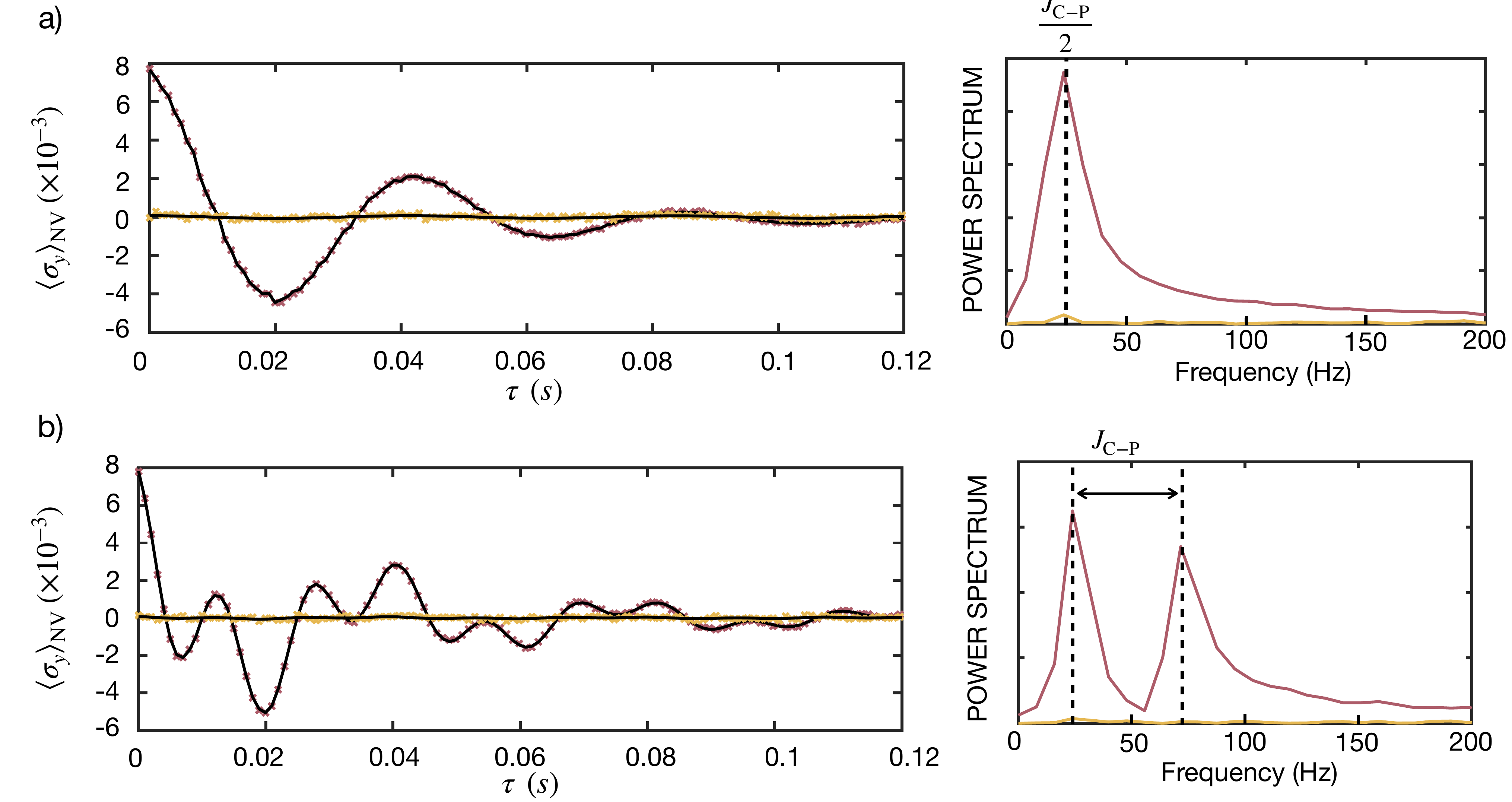}
\caption{ $\langle\sigma_{y}\rangle_{\rm NV}$  (left) and its Fourier transform spectra (right) for the $\rm{P\left(CH_3\right)_3}$ molecule. Red data correspond to our proposed method, while yellow to the standard approach via interrogation of $^{13}$C, along with their respective fits (solid black lines). The total experimental time is \( t_{\rm tot} = 10^3 \) s for both cases. Again, simulations account for NV fluorescence with a 7\% contrast. 
\textbf{(a)} Both sequences include $\pi$-pulses during the \textit{loading} stages. Obtained spectra show peaks at the same frequency, but our method results in a higher SNR.
\textbf{(b)} Midway $\pi$-pulses are removed during the \textit{loading} stages for chemical shift detection. Resonance peaks are now center at the chemical shift value, with a splitting owing  to the J-coupling between carbon and nitrogen.
}
\label{fig:PCH33_results}
\end{figure*}

In the \textit{transfer} stage \circled{2}, we apply $U_1$ leading to
\begin{align}
\rho_3\rightarrow&-B_H\sum_{i=1}^{N_H}S_i^xe^{-2i\left(\sum\limits_{r=1}^{N_t}J_{i,r}^{\rm het}\:S_i^y\:I_r^x\right)t}\nonumber\\[8pt]
&\times \left[\sum_{j=1}^{N_t}A_j^i\:I_j^xe^{-2i\:\left(\sum\limits_{k=1}^{N}J_{j,k}I_j^y\:\sigma_k^z\right)\:\tau}e^{2i\:\delta_j\:I_j^z\tau} +\sum_{j=1}^{N_t}\sum_{p=1}^{N_t}\sum_{q=1}^{N_t}A_{j,p,q}^i\:I_j^x\:I_p^x\:I_q^xe^{-2i\sum\limits_{k=1}^{N}\left(J_{j,k}I_j^y+J_{p,k}I_p^y+J_{q,k}I_q^y\right)\sigma_k^z\:\tau}e^{2i\left(\:\delta_j\:I_j^z+\delta_p\:I_p^z+\delta_q\:I_q^z\right)\tau}+...\right]\nonumber\\[8pt]
&\times e^{-2i\left(\sum\limits_{r=1}^{N_t}J_{i,r}^{\rm het}\:S_i^y\:I_r^x\right)t},
\label{eq:app:rho_3}
\end{align}

Since only $S_i^z$ are interrogated, the cosines of the exponential term $e^{-2i\left(\sum\limits_{k=1}^{N}J_{j,k}I_j^y\:\sigma_k^z\right)\:\tau}e^{2i\:\delta_j\:I_j^z\tau}$ contribute to the signal

\begin{align}
&-B_H\sum_{i=1}^{N_H}S_i^xe^{-2i\left(\sum\limits_{r=1}^{N_t}J_{i,r}^{\rm het}\:S_i^y\:I_r^x\right)t}\sum_{j=1}^{N_t}-A_j^i\:I_j^x\prod_k^{N}\left[\cos{\left(\frac{J_{j,k}}{2}\tau\right)}\right]\cos{\left(\delta_j\tau\right)} e^{-2i\:\left(\sum\limits_{r=1}^{N_t}J_{i,r}^{\rm het}\:S_i^y\:I_r^x\right)t},
\label{eq:app:sign_1}
\end{align}

where only the term with a sine function and $r=j$ in the remaining exponential survive. Then, 

\begin{align}
&+B_H\sum_{i=1}^{N_H}S_i^z\sum_{j=1}^{N_t}\left\{\prod_k^{N}\left[\cos{\left(\frac{J_{j,k}}{2}\tau\right)}\right]\sin^2{\left(\frac{J_{i,j}^{\rm het}}{2}t\right)}\prod_{k\neq j}^{N_t}\left[\cos^2{\left(\frac{J_{i,k}^{\rm het}}{2}t\right)}\right]\right\} ,
\label{eq:app:sign_1}
\end{align}
We optimize the time $t$ such the term $\sin^2{\left(\frac{J_{i,j}^{\rm het}}{2}t\right)}\prod\limits_{k\neq j}^{N_t}\left[\cos^2{\left(\frac{J_{i,k}}{2}t\right)}\right]$ is maximal.

Following the same methodology, one obtains for the second term in Eq.~(\ref{eq:app:rho_3})

\begin{align}
&B_H\sum_{i=1}^{N_H}S_i^z\sum_{j=1}^{N_t}\sum_{p=1}^{N_t}\sum_{q=1}^{N_t}\left\{\prod_k^{N}\left[\cos{\left(\frac{J_{j,k}}{2}\tau\right)}\:\cos{\left(\frac{J_{p,k}}{2}\tau\right)}\:\cos{\left(\frac{J_{j,k}}{2}\tau\right)}\right]\sin^2{\left(\frac{J_{i,j}^{\rm het}}{2}t\right)}\:\sin^2{\left(\frac{J_{i,p}^{\rm het}}{2}t\right)}\:\sin^2{\left(\frac{J_{i,q}^{\rm het}}{2}t\right)}\prod_{\substack{k\neq j\\k\neq p\\k\neq q}}^{N_t}\left[\cos^2{\left(\frac{J_{i,k}^{\rm het}}{2}t\right)}\right]\right\} ,
\label{eq:app:sign_2}
\end{align}

Note that if $t$ has been optimized to maximize the term in Eq.~(\ref{eq:app:sign_1}), this term---and the subsequent ones---is generally much smaller in most cases. Only in situations where the molecule exhibits high symmetry these terms significantly contribute to the signal. 

In Fig.~\ref{fig:PCH33_results}, we present the numerical simulation results for the $\rm{P(CH_3)_3}$ molecule discussed in the main text, comparing our sequence (red data) with the standard approach (yellow data). Fig.~\ref{fig:PCH33_results}a) shows the results with the midway $\pi$-pulses retained, while Fig.~\ref{fig:PCH33_results}b) illustrates the results with the $\pi$-pulses removed, allowing the accumulation of information due to the chemical shift.

\section{From the sample to the NV}\label{app:readout}

In order to detect the target quantities, the state of hydrogen nuclei  measured with the NV-ensemble through an induced NMR signal. As depicted in the main text, Fig.~\ref{fig:sequence} a), this occurs during the detection stages, where a radio-frequency (RF) driving targeting the hydrogen nuclei is applied on the sample and, simultaneously, a XY4 MW control sequence is applied over the NV ensemble.

To reproduce the signal received by each NV, we follow \cite{meriles2010imaging}, where the nuclear spin oscillations are reinterpreted as a classical NMR signal. The resultant field emanated by the hydrogens during the detection stage in the $k$-th block is

\begin{align}
B_z(t)= B_0(k\tau)\: \gamma_H\sin(\Omega^H t)
\label{eq:app:SampleSignal}
\end{align}

where $\Omega^H=\gamma_H B_{RF}$ is the Rabi frequency of the RF driving, and  
$$B_0(k\tau)=\frac{(2\pi)^2(\hbar)^2\:\gamma_H\mu_0\:\rho_H\: B_z }{16\pi K_B T} F_3 \langle\ \frac{2}{N^H}\sum_{i=1}^{N_H}{S^i_z}\rangle (k\tau).$$ 
Here $\hbar=1.054\cdot 10^{-34}\:\rm{J\cdot s}$ is the reduced Planck constant, $\gamma_H=(2\pi)\times42.57\: \rm{MHz/T}$ is the gyromagnetic ratio of the hydrogen, $\mu_0=4\pi\cdot10^{-7}\:\rm{H/m}$ the magnetic permeability of free space, $\rho_H$ the hydrogen density of the sample, $B_z=2\:\rm{T}$ the external magnetic field, $K_B=1.38\cdot10^{-23}\:\rm{J/K}$ the Boltzmann constant, $T=300$ K, and $F_3$ characterizes the geometry and position of the sample relative to the NV ensemble \cite{meriles2010imaging}. The maximum value, $F_3 = 4.1$, is achieved when the diamond is cut so that the [111] direction is perpendicular to the surface. We take a hydrogen density of $\rho_H=6.6\cdot10^{28}\: m^{-3}$. 

The NV evolves according to
\begin{align}
H_{NV}=\frac{\sigma_z\:\gamma_e}{2}B_z(t)+H_c,
\label{eq:app:NVHamiltonian}
\end{align}
where $H_C$ represents the MW driving over the NV ensemble. In particular, applying a resonant XY4 control sequence yields 
\begin{align}
    \langle\sigma_y\rangle_{NV}\approx\frac{\gamma_e\:\gamma_H}{\Omega^H}B_0(k\tau),
    \label{eq:app:NV signal}
\end{align}
provided that $\frac{\gamma_e\gamma_H }{\Omega^H}\:B_0(n\tau)<<1$. 

In these calculations, the decoherence time of the NV ensemble, $T_2^{\rm NV}$, has not been considered. In fact, it can be on the order of (or shorter than) the required  $2\pi$ and $-2\pi$ rotations (see main text), which significantly reduces the amplitude of the accumulated phase. In terms of sensitivity, the optimal detection time is $T_2^{\rm NV}$\cite{barry2020sensitivity}. Thus, we set such time for the measurement, but the keep the $2\pi$ and $-2\pi$ rotations over the sample, of duration $t_{\rm RF}$, for robustness purposes. With this modification, the expected value is

\begin{align}
    \langle\sigma_y\rangle_{NV}\approx\frac{2\gamma_e\:\gamma_H\:}{\Omega^H}B_0(k\tau)\left[1-\cos{\left(\pi\:T_2^{\rm NV}\Omega^H\right)}\right],
    \label{eq:app:NVsignal_time}
\end{align}

Targeting a distinct nucleus, i.e. $I_1$, one has to change $\gamma_H$ and $\Omega^H$ for $\gamma_1$ and $\Omega^1$ in the previous expressions. 

\begin{figure}[]
\includegraphics[width=1 \linewidth]{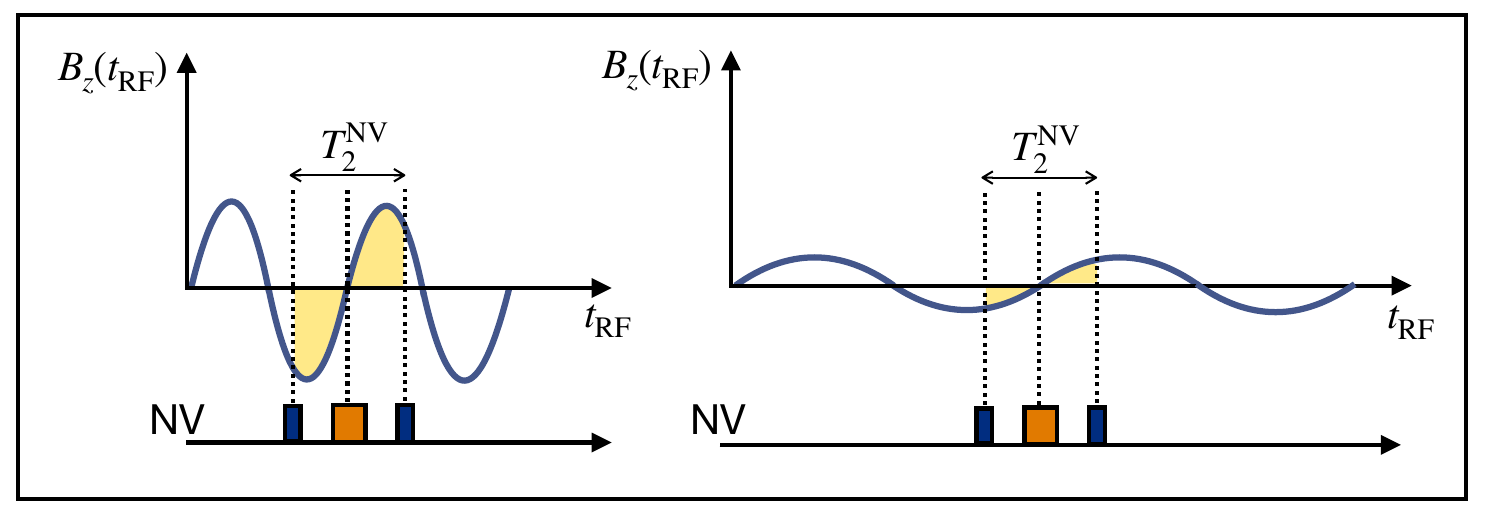}
\caption{Detection by the NV ensemble through an echo of the magnetic field induced by the sample. When rotation is induced in a nucleus with a high gyromagnetic ratio (e.g., H), both the amplitude and frequency of the generated magnetic field (Eq.~\ref{eq:app:SampleSignal}) are greater (left) compared to a nucleus with a lower gyromagnetic ratio (right). Consequently, the accumulated phase in the NV ensemble (proportional to the shaded yellow area) is higher in the high gyromagnetic ratio case, as described by Eq.~\ref{eq:app:NV signal}.}
\label{fig:nvreadout}
\end{figure}

Fig.~\ref{fig:nvreadout} depicts schematically this idea of the readout. Notice how the accumulated phase of the NV ensemble is higher for higher gyromagnetic ratios (Fig.~\ref{fig:nvreadout} left) compared to nucleus with lower gyroagnetic ratios (Fig.~\ref{fig:nvreadout} right), as one can see from Eq.~\ref{eq:app:NVsignal_time}.

\section{Standard approach}\label{app:simpler}

Here, we describe the standard approach for detecting the quantities of interest related to $I_1$~\cite{alsina2024jcoupling}. The pulse sequence is shown in Fig.~\ref{fig:sequence} c). Note that we consider a prepolarization stage is applied so that the initial polarization of nucleus $I_1$ matches that of the hydrogen. During the \textit{loading} stage \circled{1} in the $k$-th block, the target quantities get encoded in the state of nucleus $I_1$, leading to

\begin{align}
	\rho_1(\tau)=\frac{1}{2^N}\left[B_1\:S_1^z+B_H\:I_1^z\:C(k\tau)+\sum_{j=2}^N B_j\:I_j^z\right],
\end{align}
with $C(k\tau)=\prod\limits_{j=2}^N\left[\cos{\left(\frac{J_{1,j}}{2}k\tau\right)}\right]$ or $C(k\tau)=\prod\limits_{j=2}^N\left[\cos{\left(\frac{J_{1,j}}{2}k\tau\right)}\right]\:\cos{\left(\delta_1k\tau\right)}$.

 Note that the RF is now applied to $I_1$ during the detection stage \circled{2}, and thus, it is $I_1$ that emits the magnetic field, which is then tracked by the NV ensemble (see right panel in Fig.~\ref{fig:nvreadout}), leading to 
 
 \begin{equation}\label{eq:app:effectiveB1}
B_z(t)\propto \gamma_1\langle I_1^z \rangle(k\tau)\sin(\Omega^\text{1} \:t),
\end{equation}
where, importantly, the amplitude of such magnetic field is now $\propto \gamma_1\langle I_1^z \rangle(k\tau)$, and the frequency $\Omega^1\propto\gamma_1$. Here,

\begin{align}
    \langle I_1^z\rangle(k\tau)=-\frac{1}{2}B_H\:C(k\tau).
    \label{eq:app:expectn_i}
\end{align}

\section{sensitivity analysis}\label{app:sensitivity}

We compute the sensitivity provided by the NV ensemble measurement in our proposal. Each block of the sequence consists of stages \circled{1}, \circled{2}, \circled{3}, and \circled{4}, and the total block duration is given by $\tau + 2t + M\cdot t_H$. Here, $\tau$ is the \textit{loading} stage time, $t$ is the \textit {transfer} stage time, $t_H$ is the irradiation time of the hydrogens, and $M$ is the number of measurements in the detection stage, see Fig.~\ref{fig:sequence}~a). During the \textit{loading} stage, the nucleus $I_1$ is subjected to dephasing, while during the remaining stages it is the hydrogen nucleus that suffers dephasing.
For each block, this decoherence  contributes to a signal attenuation factor
$$e^{-\left(\frac{2t + M\cdot t_H}{T_2^H} + \frac{\tau}{T_2^1}\right)},$$
where $T_2^H$ is the hydrogen decoherence time, and $T_2^1$ is the decoherence time of nucleus $I_1$.

Since the spectroscopically useful time is $\tau$, let us define the following effective decoherence time, $T_{2,\rm eff}^{H}=T_2^H\:T_2^1\:\frac{\tau}{\left(2\:t+M\cdot t_H\right)T_2^1+\tau\:T_2^H}$, so that the attenuation factor can be written as
$$e^{\frac{\tau}{T_{\rm 2, eff}^H}}.$$

For an oscillatory signal with total duration $n\tau$ (where $n$ is the total number of applied blocks in our sequence) and an attenuation factor as described earlier, one can show that the height of the frequency peak in the Fourier spectrum will scale as
$$\propto T_{2,\rm eff}^H\left(1 - e^{-\frac{\tau}{T_{2, \rm eff}^H}n}\right).$$

Taking this into account, the sensitivity provided by the NV ensemble measurement in our proposal, $\eta_{\rm NV}^H$, assuming an optimal measurement time of $T_2^{\rm NV}$ (see Appendix~\ref{app:readout} and \cite{barry2020sensitivity}).

\begin{align}
    \eta_{\rm NV}^H\propto \frac{1}{T_{2,\rm eff}^H\left(1-e^{-\frac{\tau}{T_{2, \rm eff}^H}n}\right)}\:\frac{e}{A_H}\:\frac{1}{\sqrt{M}}\:\frac{1}{\sqrt{V}},
    \label{eq:sensitivity_general_H}
\end{align}

where $A_H$ is the term corresponding to the accumulated phase by the NV ensemble during the readout stage, Eq.~(\ref{eq:app:NVsignal_time}), and $V$ corresponds to the number of experimental averages. 
Writing $V$ in terms of the experimental time, $V=\frac{t^{\rm exp}}{n\:\left(2\:t+\tau+M\cdot t_H\right)}$, the new expression for the sensitivity reads

\begin{align}
    \eta_{\rm NV}^H\propto  \frac{1}{T_{2,\rm eff}^H\left(1-e^{-\frac{\tau}{T_{2, \rm eff}^H}n}\right)}\:\frac{e}{A_H}\:\sqrt{\frac{n\left(2\:t+\tau+M\cdot t_H\right)}{t^{\rm exp}}}\:\frac{1}{\sqrt{M}},
    \label{eq:sensitivity_general_H_opt}
\end{align}

For the simpler sequence (where we denote the target nucleus as $I_1$), we find a similar expression

\begin{align}
    \eta_{\rm NV}^1\propto \frac{1}{T_{2,\rm eff}^1\left(1-e^{-\frac{\tau}{T_{2, \rm eff}^1}n_1}\right)}\:\frac{e}{A_1}\:\sqrt{\frac{n_1\left(\tau_1+M_1\cdot t_1\right)}{t^{\rm exp}}}\:\frac{1}{\sqrt{M_1}},
     \label{eq:sensitivity_general_N_opt}
\end{align}
where the effective decoherence time is here $T_{2,\rm eff}^{1}=T_2^1\:\frac{\tau}{\tau+M_1\cdot t_1}$.

The proportionality constant in Eq.~(\ref{eq:sensitivity_general_H_opt}) and Eq.~(\ref{eq:sensitivity_general_N_opt}) is the same.

The fairest comparison between both methods would be to set the loading periods and number of measurements to have the same duration, $\tau_1=\tau$ and $n_1=n$. With such assumption, the sensitivity ratio between both methods is

\begin{align}
    \frac{\eta_{\rm NV}^H}{\eta_{\rm NV}^{1}}&=\left[\frac{1-e^{-\frac{\tau}{T_{2, \rm eff}^1}n}}{1-e^{-\frac{\tau}{T_{2, \rm eff}^H}n}}\right]\:\frac{T_{2,\rm eff}^1}{T_{2,\rm eff}^H}\: \frac{A_1}{A_H}\sqrt{\frac{2\:t+\tau+M\cdot t_H}{\tau+M_1\cdot t_1}}\sqrt{\frac{M_1}{M}}.
    \label{eq:sensitivity_general}
\end{align}

\end{document}